\documentclass[aps, prd, amsmath, amssymb, nofootinbib, preprint]{revtex4-1}

\usepackage[dvipsnames]{xcolor} % Expand color palette
\usepackage[hyperindex, breaklinks, colorlinks=true, linkcolor={violet}, citecolor={OliveGreen}, urlcolor={NavyBlue}]{hyperref} % hyperref and hyperref setup 
\usepackage{bm} % Boldsymbol
\usepackage{tensor} % Tensor indices

\newcommand{\ind}[1]{\indices{#1}} % redefining `indices' command from tensor package
\newcommand{\Q}{\ensuremath{\mathcal{Q}}} % Radial equation One
\newcommand{\R}{\ensuremath{\mathcal{R}}} % Radial equation Two
\newcommand{\M}{\ensuremath{\mathcal{M}}} % spacetime manifold
\renewcommand{\O}{\ensuremath{\mathcal{O}}} % Big-O
\newcommand{\eref}[1]{Eq.~(\ref{#1})} % equation reference (type one)
\newcommand{\erefs}[2]{Eqs.~(\ref{#1}) and (\ref{#2})} % equation reference (type three)
\newcommand{\rref}[1]{Ref.~\cite{#1}} % reference reference
\newcommand{\defeq}{\equiv} % Definition symbol
\renewcommand{\d}{\text{d}} % Differential symbol
\renewcommand{\vec}[1]{\boldsymbol{#1}} % Vector (bold) symbol

\begin{document}

\title{Electric Multipole Fields of Higher-Dimensional Massive Bodies}
\author{Matthew S. Fox}
\email{msfox@g.hmc.edu}
\affiliation{Department of Physics, Harvey Mudd College, Claremont, CA 91711, USA}
\date{\today}

\begin{abstract}
It was shown in a recent paper [\href{https://doi.org/10.1063/1.5124502}{\emph{J. Math. Phys.} \textbf{60}, 102502 (2019)}] that slowly lowering an electric charge into a Schwarzschild-Tangherlini (ST) black hole endows the final state with electric multipole fields, which implies the final state geometry is not Reissner-Nordstr\"om-Tangherlini in nature. This conclusion departs from the four-dimensional case in which the no-hair theorem (NHT) requires the final state to be a Reissner-Nordstr\"om black hole. To better understand this discrepancy clearly requires a deeper understanding of the origin of the multipole hair in the higher-dimensional case. In this paper, we advance the conjecture that charged, static, and asymptotically-flat higher-dimensional black holes can acquire electric multipole hair only after they form. This supposition derives from studying the asymptotic behavior of the field of a multipole charge onto which a massive and hyperspherical shell with an exterior ST geometry is collapsing. In the mathematical limit as the shell approaches its ST radius, we find that the multipole fields (except the monopole) vanish. This implies that the only information of an arbitrary (but finite) charge distribution inside the collapsing shell that is available to an asymptotic observer is the total electric charge. Our results yield considerable insight into how higher-dimensional black holes acquire electric multipole hair, and also imply that, in four dimensions, the fadeaway of multipole moments during gravitational collapse is not strictly because of the NHT.
\end{abstract}

\maketitle

\section{Introduction}
\label{sec:Intro}

Extra spatial dimensions are now a precondition for consistency in many approaches to quantum gravity (e.g., string theory). Furthermore, the AdS/CFT correspondence relates the properties of an $n + 1$-dimensional black hole to those of an $n$-dimensional quantum field theory \cite{Maldacena, Maldacena2}. For these (and other \cite{Emparan1}) reasons, it is imperative in string theory, and other approaches to quantum gravity, to have a keen intuition for how higher-dimensional black holes behave. In addition, as we illustrate below, studies into higher-dimensional black holes can yield insights into the character of well-known features of four-dimensional black holes, which only bolsters our understanding of them.

Consider first four-dimensional spacetime. Here, black holes are stringently constrained by Wheeler's no-hair theorem (NHT) \cite{Wheeler}, which states that all four-dimensional, stationary, and asymptotically-flat black hole solutions to the Einstein-Maxwell equations are completely characterized by just three independent parameters: mass, angular momentum, and electric charge \cite{Israel1, Israel2, Carter1, Carter2, Hawking, Robinson, Heusler}. This theorem enables us to straightforwardly predict the final state of a static black hole that is subjected to a slow\footnote{By ``slow'' we mean ``slow enough that the static considerations remain valid.''} physical process.

Consider, for example, slowly lowering an electric charge of strength $q$ into a Schwarzschild black hole of mass $M$. Evidently, the final state will be a static black hole with mass $M$ and charge $q$. However, it is not immediately clear if, in addition, the final state will possess unconserved charges like electric multipole moments (excluding the monopole). Rest assured, in order that it not have such multipole hair, the NHT requires the final state geometry to be the spherically-symmetric Reissner-Nordstr\"om (RN) solution. Indeed, this agrees with the result of the more detailed analysis in \rref{Cohen}. Thus, even though the charge distribution is highly asymmetrical, the electrostatic potential approaches that of the spherically-symmetric RN black hole as the charge nears the horizon.

The story is strikingly different in higher-dimensional spacetimes. Here, black holes are considerably less constrained than four-dimensional ones, largely for two reasons.\footnote{See \rref{Emparan0} for a separate and less heuristic perspective.} First, there are more rotational degrees of freedom in an $n + 1$-dimensional spacetime, which means stationary black holes become progressively more complex as $n$ increases \cite{Hollands, Emparan1}. Moreover, if $n \geq 5$, then black holes with fixed masses can have arbitrarily large angular momentum \cite{Myers}. Second, Hawking's topology theorem \cite{Hawking} (a subtle piece of the proof of the NHT) fails because it relies on the Gauss-Bonnett theorem. This implies the boundary topology of an $n + 1$-dimensional black hole need not be homeomorphic to the $n-1$-sphere. Of course, topological restrictions do exist when $n > 3$ \cite{Galloway1, Galloway2, Helfgott}, but more than one boundary topology is allowed \cite{Emparan1, Emparan2, Costa}.

These results imply that the uniqueness theorems for four-dimensional black holes do not readily generalize to higher dimensions. Though uniqueness theorems of static, higher-dimensional black holes exist, in proving them you must include the additional assumption of a nondegenerate horizon (a property you get for free when $n = 3$ \cite{Chrusciel1, Chrusciel2}) \cite{Hollands, Hwang, Rogatko1, Rogatko2, Gibbons1, Gibbons2, Gibbons3, Ida}. Nevertheless, once restricted to solutions with regular horizons, the natural dimensional continuations of the well-known $n = 3$ solutions emerge. For example, the Schwarzschild-Tangherlini (ST) black hole is the unique static and asymptotically-flat vacuum solution to the higher-dimensional Einstein equations \cite{Hollands, Hwang, Gibbons1, Gibbons2}.\footnote{We use the term ``asymptotically-flat'' in the sense used in the higher-dimensional uniqueness theorems. See \rref{Hollands} for the relevant rigorous definitions.} It is therefore the natural extension of the Schwarzschild black hole to higher dimensions \cite{Tangherlini}. Similarly, the Reissner-Nordstr\"om-Tangherlini (RNT) black hole is the unique static and asymptotically-flat electrovac solution to the higher-dimensional Einstein-Maxwell equations \cite{Gibbons3, Ida}, making it the natural generalization of the Reissner-Nordstr\"om solution to higher dimensions \cite{Tangherlini}.

Given this parallel between the unique $n = 3$ and $n \neq 3$ static solutions, one may expect the behavior of the $n \neq 3$ solutions to mimic that of the $n = 3$ solutions when subjected to an identical physics process (albeit in a higher dimension). This, however, is not correct, as the previous example with the electric charge will show.

Consider the same electric charge $q$ from before, but this time slowly lower it into an ST black hole with mass $M$. Again, the final state is a static black hole with mass $M$ and charge $q$. However, due to the weaker assumptions underlying the higher-dimensional uniqueness theorems, in order to conclude that the final state geometry is RNT in nature, one needs to also show that this process does not affect the regularity of the horizon. Surprisingly, as shown in \rref{Fox}, this or the horizon topology is compromised during the infall of the charge,\footnote{This assumes the spatial dimension $n$ is odd. If $n$ is even, then the energy density of the electric field diverges as the charge approaches the horizon, which imposes unbounded stresses on the horizon and leads to an apparent violation of asymptotic flatness \cite{Fox}. In either case, an RNT black hole is not produced.} which means that the final state is \emph{not} RNT in nature. Ultimately, these conclusions follow from the fact that the infalling charge furnishes the final state black hole with electric multipole hair.

This simple example illustrates a profound difference in the response of $n = 3$ and $n \neq 3$ black holes to a straightforward physical process. Whereas the multipole fields of the charge vanish as the charge approaches the event horizon of the four-dimensional Schwarzschild black hole, they do not as the charge approaches the horizon of the higher-dimensional ST black hole. Clearly, to better understand this discrepancy requires a deeper understanding of the origin of multipole hair on higher-dimensional black holes. To this end, we study in this paper the plausibility of a static, hyperspherical, and asymptotically-flat higher-dimensional black hole forming with multipole hair. Can a higher-dimensional black hole form with multipole hair? Or must it be acquired by infalling electric charges after the black hole forms?

In four dimensions, Wald explicitly showed that the collapse of a spherical and massive shell onto a finite distribution of electric multipole charges completely suppresses the multipole fields (except the monopole) \cite{Wald}. This, of course, agrees with the NHT, and suggests that a four-dimensional black hole cannot form with electric multipole moments.

To simulate the formation of a static, higher-dimensional black hole, we employ the obvious generalization of Wald's setup to higher dimensions in which, in the exterior spacetime region, the collapsing shell has an ST geometry. By placing a multipole charge at the center of the shell, we are able to examine the response of the asymptotic multipole field to the inward collapse of the shell. Like Wald, we model this collapse as a sequence of static shell solutions converging to their common ST radius (the higher-dimensional Schwarzschild radius). In this limit, we find that the multipole fields are completely suppressed (except the monopole). This implies that the only information of an arbitrary (but finite) charge distribution inside a collapsing, higher-dimensional shell that can be measured by a distant observer is the total electric charge. Based on our calculations, we conjecture that charged, static, and asymptotically-flat higher-dimensional black holes can acquire electric multipole hair only after they form. This affords considerable insight into how higher-dimensional black holes acquire electric multipole moments: charges must fall into them \emph{after} formation.

\section{Multipole Field Suppression via Higher-Dimensional Black Hole Formation}
\label{sec:ShellCollapse}

The ST spacetime metric (and the shell metric below) is most naturally expressed in ST coordinates $\vec{\psi} = (t,r,\vec{\varphi})$, where $\vec{\varphi} = (\varphi_1, \dots, \varphi_{n-1})$ are the standard hyperspherical coordinates on the unit $n - 1$-sphere. As in the Schwarzschild case, $t$ is interpreted physically as ``time to an asymptotic observer'' and $r$ as ``circumferential radius to an asymptotic observer.'' 

The spacetime metrics of various higher-dimensional shells have been studied in models of higher-dimensional gravitational collapse. See, e.g., \rref{Mena} and references therein for a rigorous overview on building such metrics, and \rref{Gao} for an insightful example into a charged shell. Ultimately, these metrics are derived in the standard way using Israel's geometric theory of spacetime junctions \cite{IsraelBoundary}. Below, we briefly summarize how this theory applies to our study.

Let $(\M, g)$ be an $n + 1$-dimensional spacetime and $\Sigma \subset \M$ a codimension-one timelike hypersurface that is to represent the shell. The problem is to determine $g$ subject to Einstein's equations and the constraints of the shell (in our case: infinitesimally-thin, massive, static, and hyperspherical). Evidently, $\Sigma$ separates $(\M,g )$ into disjoint ``exterior'' and ``interior'' spacetimes, denoted by $(\M^+, g^+)$ and $(\M^-, g^-)$, respectively. Both of these spacetimes have a boundary diffeomorphic to $\Sigma$, which allows one to relate the local coordinates in the exterior region to the local coordinates in the interior region via the coordinates on the shell \cite{Mena}. In this paper, we choose the exterior region $(\M^+, g^+)$ to be ST spacetime and the interior region $(\M^-, g^-)$ to be Minkowski spacetime. These choices fix the exterior and interior metrics $g^+$ and $g^-$, respectively, which can then be expressed in terms of two sets of ST coordinates $\vec{\psi}^+$ and $\vec{\psi}^-$. The remaining task is to relate $\vec{\psi}^+$ and $\vec{\psi}^-$ using the jump conditions across $\Sigma$ \cite{IsraelBoundary}. In our case, this amounts to integrating the field equation $G\ind{^0_0} = 8\pi T\ind{^0_0}$ in local coordinates over a ``pillbox'' on $\Sigma$ \cite{MTW}. The result is
\begin{equation}
g(\d \vec{\psi}, \d \vec{\psi}) = 
\begin{cases}
-\left(1 - \frac{r_s^{n-2}}{R^{n-2}} \right)\d t^2 + \d r^2 + r^2 \gamma(\d\vec{\varphi},\d\vec{\varphi}), & r < R,\\
-\left(1 - \frac{r_s^{n-2}}{r^{n-2}} \right)\d t^2 + \left(1 - \frac{r_s^{n-2}}{r^{n-2}} \right)^{-1}\d r^2 + r^2 \gamma(\d\vec{\varphi},\d\vec{\varphi}), & r > R,
\end{cases}
\label{eq:metric}
\end{equation}
where $R$ is the radius of the shell, $r_s$ is the ST radius,\footnote{To ensure the Minkowskian interior of the shell, we assume the mass of the shell is such that $r_s < R$.} and $\gamma$ is the standard metric on the unit $n-1$-sphere. Since \eref{eq:metric} with $n = 3$ reduces to the spacetime metric used by Wald in \rref{Wald}, our model is indeed a higher-dimensional generalization of that study.

We now calculate the field of an electrostatic $k$-pole of strength $\sigma_k$ placed at the center ($r = 0$) of the hyperspherical shell. We assume $\sigma_k$ is small enough that its influence on the background geometry is negligible. Under this condition and that of electrostaticity, the Faraday two-form $F$ is simply $F = \d(\Psi \d t)$, where, within the shell, the scalar field $\Psi$ satisfies the source-free Maxwell equations in an $n + 1$-dimensional Minkowski spacetime,
\begin{equation}
\Psi(r,\vec{\varphi}) = \left[a_k r^{k} + b_k r^{-(k + n - 2)}\right]Y_k(\vec{\varphi}), \quad r < R.
\label{eq:freeFieldSolutionInside}
\end{equation}
For a $k$-pole of strength $\sigma_k$ at $r = 0$,
\begin{equation}
b_k = \sqrt{1 - \left(\frac{r_s}{R}\right)^{n-2}}\, \sigma_k,
\label{eq:multipoleCoefficient}
\end{equation}
where the square-root factor follows from the conversion of coordinate time to proper time inside the shell when calculating the orthonormal frame components of $F_{\mu\nu}$. Incidentally, in \eref{eq:freeFieldSolutionInside} we are denoting by $Y_k(\vec{\varphi})$ the sum over all orders of the degree $k$ hyperspherical harmonic functions. However, the details of these functions (see \rref{Frye}) are immaterial for this analysis because the (infinitesimally-thin) shell is hyperspherically-symmetric around the multipole charge, so the angular fields $Y_k(\vec{\varphi})$ are insensitive to the shell.

Outside the shell, $\Psi$ satisfies the source-free Maxwell equations in an $n + 1$-dimensional ST spacetime \cite{Fox},
\begin{equation}
\Psi(r, \vec{\varphi}) = \left[c_k \Q_k(r) + d_k\R_k(r)\right] Y_k(\vec{\varphi}), \quad r > R,
\label{eq:freeFieldSolutionOutside}
\end{equation}
where $\Q_k$ and $\R_k$ are the hypergeometric series
\begin{align}
\Q_k(r) &= r_s^{-(k + n - 2)} \sum_{m \geq 0} \frac{\left(1 + \frac{k}{n-2}\right)_m \left(\frac{k}{n-2}\right)_m}{m!\left(2 + \frac{2k}{n-2}\right)_m}\left(\frac{r_s}{r}\right)^{k + (m+1)(n-2)},
\label{eq:seriesSolutionOneComplete}\\
\R_k(r) &= r_s^{k} \sum_{m = 0}^{\Lambda_{k}} \frac{\left(-\frac{k}{n-2}\right)_m \left(-1 - \frac{k}{n-2}\right)_m}{m!\left(-2 - \frac{2k}{n-2}\right)_m} \left(\frac{r}{r_s}\right)^{k - m(n-2)}.
\label{eq:seriesSolutionTwoComplete}
\end{align}
Here, $(x)_m \equiv x(x+1)\cdots(x + m - 1)$ is the Pochhammer symbol, defined such that $(x)_0 = 1$ for all real $x$. The summation bound $\Lambda_k$ in \eref{eq:seriesSolutionTwoComplete} derives from an elementary number-theoretic relation between the moment $k$ of the multipole charge $\sigma_k$ and the dimensionality $n$ of the space. However, the precise details (see \rref{Fox}) are again unimportant because the requirement of regularity of $\Psi$ as $r\rightarrow \infty$ implies $d_k = 0$, so the $\R_k$ solution leaves the analysis entirely.

We can now determine the remaining coefficients $a_k$ and $c_k$ in \erefs{eq:freeFieldSolutionInside}{eq:freeFieldSolutionOutside}, respectively, via the jump continuity constraints on $F$ across the $r = R$ boundary, i.e., the requirement that the orthonormal frame components of $F_{\mu\nu}$ be continuous across the shell. These are
\begin{equation}
\lim_{r \rightarrow R^+} \Psi(r, \vec{\varphi}) = \lim_{r \rightarrow R^-} \Psi(r, \vec{\varphi})
\label{eq:conditionOne}
\end{equation}
and, by the assumption that the boundary itself is electrically-neutral,
\begin{equation}
\lim_{r \rightarrow R^+} \partial_r\Psi(r, \vec{\varphi}) = \lim_{r \rightarrow R^-} \frac{\partial_r\Psi(r, \vec{\varphi})}{\sqrt{1 - \frac{r_s^{n-2}}{R^{n-2}}}}.
\label{eq:conditionTwo}
\end{equation}
Together, \erefs{eq:conditionOne}{eq:conditionTwo} imply
\begin{align}
a_k &= \frac{\alpha \left[\alpha R + \Q_k(k + n - 2)\right]\sigma_k}{(k\Q_k - \Q'_k \alpha R)R^{2k + n - 2}},
\label{eq:aEquation}\\
c_k &= \frac{\alpha(2k+n-2)\sigma_k}{\left(k\Q_k - \Q_k'\alpha R\right)R^{k+n-2}},
\label{eq:cEquation}
\end{align}
where $\alpha(R) \defeq \sqrt{1 - (r_s / R)^{n-2}}$ and a prime denotes a derivative with respect to $r$. For sake of clarity, we have dropped the argument of the shell radius $R$ when writing $c_k, \Q_k, \Q_k'$, and $\alpha$ in \erefs{eq:aEquation}{eq:cEquation}, and we shall adopt this convention hereafter. Therefore, unless explicitly stated, $c_k, \Q_k$, $\Q'_k$, and $\alpha$ are implicitly evaluated at the radius of the shell for the remainder of this article.

Now, it is evident from \eref{eq:seriesSolutionOneComplete} that, asymptotically,
\begin{equation}
\Q_k(r) \sim r^{-(k + n - 2)} \left[1 + \O\left(\frac{r_s}{r}\right)\right].
\label{eq:bigO}
\end{equation}
Hence, $c_k$ is the electrostatic $k$-pole moment measured by a distant observer when the shell radius is $R$. 

For the monopole case ($k = 0$), a distant observer measures $c_0 = \sigma_0$ because $\Q_0 = 1 / R^{n-2}$ [see \eref{eq:seriesSolutionOneComplete}]. In words, a massive and hyperspherical shell does not disrupt the field of an electrostatic monopole charge, as one would expect. If $k \neq 0$ (i.e., $k > 0$), then $c_k \neq \sigma_k$. However, if $R \gg r_s$, then $\Q_k \approx R^{-(k + n - 2)}$ by \eref{eq:bigO}, so a distant observer measures $c_k \approx \sigma_k$ by \eref{eq:cEquation}. This implies that a massive and hyperspherical shell that is considerably larger than its own ST radius only weakly disrupts the moment of an electrostatic multipole charge contained inside it.

Evaluating the opposite limit, where the shell radius $R$ approaches the ST radius $r_s$, is less straightforward. Of course, this limit makes physical sense if and only if $R$ approaches $r_s$ from above ($R \rightarrow r_s^+$), so the precise mathematical problem is to evaluate $c_k$ as $R\rightarrow r_s^+$ when $k \neq 0$. We shall prove the limit vanishes, which means that the field of the multipole charge does not escape the resulting black hole. To do this, we introduce the coordinate $\rho(R) \defeq (r_s / R)^{n-2}$, in terms of which $c_k$ maps to the function
\begin{equation}
c_k(\rho) = \frac{\sqrt{1-\rho}}{I_k(\rho)\, _2F_1(\lambda_k + 1, \lambda_k, 2\lambda_k + 2; \rho) + J_k(\rho) \, _2F_1(\lambda_k + 2, \lambda_k + 1, 2\lambda_k + 3; \rho)\sqrt{1-\rho}},
\label{eq:functionalForm}
\end{equation}
where $_2F_1$ is Gauss' hypergeometric function and $\lambda_k, I_k,$ and $J_k$ are the following real-valued expressions:
\begin{align}
\lambda_k &= \frac{k}{n-2},\\
I_k(\rho) &= k + (n-2)(1 + \lambda_k)\sqrt{1-\rho},\\
J_k(\rho) &= \frac{k\rho}{2}.
\end{align}
We seek the limit of $c_k(\rho)$ as $\rho \rightarrow 1^-$ when $k \neq 0$. Using Euler's integral representation of $_2F_1$,
\begin{equation}
_2F_1(a,b,c; \rho) = \frac{\Gamma(c)}{\Gamma(b)\Gamma(c - b)} \int_0^1 \frac{t^{b-1}(1-t)^{c-b-1}}{(1-\rho t)^{a}}\, \d t,
\label{eq:EulerIdentity}
\end{equation}
which is valid for $|\rho| < 1$ provided $b$ and $c$ are real and such that $c > b > 0$ \cite{Bailey}, it is straightforward to show that $I_k(\rho)\, _2F_1(\lambda_k + 1, \lambda_k, 2\lambda_k + 2; \rho)$ is finite and nonzero as $\rho \rightarrow 1^-$. Additionally, one can show that $_2F_1(\lambda_k + 2, \lambda_k + 1, 2\lambda_k + 3; \rho)$ has a logarithmic singularity as $\rho \rightarrow 1^- $, which implies
\begin{equation}
J_k(\rho)\, _2F_1(\lambda_k + 2, \lambda_k + 1, 2\lambda_k + 3; \rho) \sim \rho\log \left(\frac{1}{1 - \rho}\right)
\label{eq:logarithmicBehavior}
\end{equation}
for $\rho \approx 1$. Thus, as $\rho \rightarrow 1^-$, the vanishing square-root factor $\sqrt{1 - \rho}$ completely overwhelms the logarithmic divergence in \eref{eq:logarithmicBehavior}, and $c_k(\rho) \rightarrow 0$ as $\rho \rightarrow 1^-$. Accordingly,
\begin{equation}
\lim_{R \rightarrow r_s^+} c_k(R) = 0, \quad k \neq 0,
\label{eq:multipolesVanish}
\end{equation}
as claimed. In words, to a distant observer, all multipole moments inside the shell (except the monopole) fade away as the shell collapses to its own ST radius.

Now suppose the shell is filled with an arbitrary (but finite) distribution of static multipole charges. In this case, the electric field outside the charge distribution (but still inside the shell) can be represented as a superposition of the various multipole fields at the center of the shell. Our analysis shows that as $R \rightarrow r_s^+$, each of these multipole fields goes to zero, with the exception of the monopole ($k = 0$). Consequently, in the limit as the shell approaches its own ST radius, the only property of an arbitrary charge distribution inside the shell that can be measured by an asymptotic observer is the total electric charge. This conclusion is identical to that obtained by Wald in the Schwarzschild ($n = 3$) case \cite{Wald}, and is what one would naturally intuit from the NHT of four-dimensional black holes.

We acknowledge that the collapse of an infinitesimally-thin shell to its ST radius is a highly idealized and unphysical model of collapsing matter. A more realistic description is the gravitational collapse of a hyperspherical ball of fluid obeying a particular equation of state. Still, even in this more complex case, there will be a net electric field (now affected of course by the dielectric effects of the fluid) that we could in principle approximate as an arbitrary and finite distribution of electric charges contained inside the shell-like boundary of the hyperspherical ball. Of course, in general the fluid inside (and thus the charges) will not be static, but in any approximation where they are, our results suggest that the net multipole moments of the interior charges will vanish as the boundary of the hypersphere collapses inward. Consequently, it is plausible that even in this more general setting, the resulting higher-dimensional black hole will not possess multipole fields following its formation. We therefore advance the conjecture that charged, static, and asymptotically-flat higher-dimensional black holes can acquire electric multipole hair only after they form. 

Ultimately, the significance of this conjecture lies in its application to the ideas that motivated it in the first place: four-dimensional black holes and the AdS/CFT correspondence. In four dimensions, it is sometimes said (even by the author \cite{Fox}) that the fadeaway of multipole moments (electric or otherwise) during gravitational collapse occurs \emph{because} of the NHT. While technically correct, our results demonstrate that there exists a dimensionally-independent explanation for the fadeaway. This follows because we have shown that the fadeaway Wald studied in four dimensions \cite{Wald} also occurs in higher dimensions---a regime in which the NHT does not apply. Hence, there must exist a deeper, dimensionally-independent property (or set of properties) of black holes that causes the fadeaway. Of course, we may speculate as to what dimensionally-independent property (or set of properties) is responsible, however, justifying such speculation invariably requires us to prove our conjecture true, which remains an open problem.

In the context of AdS/CFT, a separate problem arises, concerning the holographic interpretation of our conjecture. While holographic interpretations of the gravitational collapse of, for example, degenerate stars exist \cite{Boer, Arsiwalla}, the author is unaware of any studies into the response of multipole moments during gravitational collapse in the context of AdS/CFT. Developing a holographic interpretation of this and our conjecture (and asymptotically de Sitter or anti de Sitter generalizations thereof) is thus an interesting avenue for future research on which we hope to report soon.

\begin{acknowledgments}
The author thanks Brennen Quigley for reviewing the present article and Cache Sanchez for many helpful discussions.
\end{acknowledgments}

\bibliography{MultipoleReferences.bib} % Bibtex bibliography

\end{document}